# Impact of snowfall on terahertz channel performance: measurement and modeling insights

Guohao Liu[1,2], Xiangkun He[3], Jiabiao Zhao[1], Da Li[1], Hong Liang[4], Houjun Sun[1,2],

Daniel M. Mittleman[5], Jianjun Ma[1,2*]

[1] School of Information and Electronics, Beijing Institute of Technology, Beijing, 100081 China
[2] Tangshan Research Institute of BIT, Tangshan, Hebei, 063099 China
[3] Department of Bioengineering, Faculty of Engineering, Imperial College London, London SW7 2AZ, UK
[4] Meteorological Observation Center of China Meteorological Administration, Beijing, 100081 China
[5] School of Engineering, Brown University, Providence, RI, 02912 USA
[*] Corresponding: jianjun_ma@bit.edu.cn

**Abstract**

In the evolving domain of wireless communication, the investigation on terahertz (THz) frequency spectrum, spanning 0.1 to 10 THz, has become a critical focus for advancing ultra-high-speed data transmission technologies. The effective deployment of THz wireless communication techniques mandates a complete study of channel performance under various atmospheric conditions, such as rain, fog, cloud, haze, and notably, snow. These environmental elements significantly impact the design of the protocol stack, ranging from physical-layer signal processing to application design and strategic network planning. An in-depth understanding of channel propagation and fading characteristics in real-world environments, especially over ultra-wide bandwidths, is crucial.

This work presents a comprehensive measurement-based and theoretical investigation of line-of-sight (LoS) THz channel performance in snowy conditions. It methodically examines both the empirical and predicted aspects of channel power and bit-error-ratio (BER). The effects of snowfall rate, carrier frequency, ambient temperature, and relative humidity on channel performance are analyzed and discussed. Our findings demonstrate that snowy conditions not only amplify power loss but also induce rapid fluctuations in the power levels of the THz channel. Notably, our results reveal an absence of significant multipath effects in these scenarios. This insight highlights the need for further research into the dynamics of snowflake movement and their interaction with THz transmission paths.

## I. Introduction

In the dynamic landscape of wireless communications, the exploration of the terahertz (THz) spectrum, encompassing frequencies from 0.1 to 10 THz, has emerged as a essential area in the pursuit of ultra-high-speed data transmission. This particular segment of the electromagnetic spectrum, notably at its lower frequency range, presents enormous potential for enabling data transmission rates exceeding 100 Gigabits per second (Gbps) [1]. Such groundbreaking capabilities are crucial to meet the burgeoning demands of various data-intensive applications, including, but not limited to, advanced cloud gaming, remote medical procedures like telesurgery, and the ever-evolving realm of the metaverse.

Despite the significant potential of THz frequencies in achieving such remarkable data rates, their practical deployment in wireless systems encounters notable challenges, predominantly due to their interaction with atmospheric elements. Contrasting with their behavior in dielectric waveguides [2, 3], THz channel propagation through free space is substantially influenced by molecular absorption, primarily from water vapor, leading to considerable signal attenuation [4-6]. This attenuation is further intensified in diverse meteorological conditions, including fog, rain, and snow [7-13]. The recent advancements in THz technology, encompassing developments in both device and radio frequency components, along with innovations in physical layer and medium access control (MAC) layer technologies, have catalyzed a trend of research dedicated to overcoming these propagation challenges. A key strategy in addressing these issues has been the deployment of high-gain directional antennas, which play a critical role in counteracting the intrinsic problems of increased spreading loss and susceptibility to obstructions, a natural characteristic of THz frequencies [14]. However, the incorporation of these advanced technologies introduces added complexities in MAC protocols and network architecture, necessitating the creation of robust, efficient, and reliable frameworks for communication links.

The effective integration of THz technology into the infrastructure of future wireless communication networks, still needs understanding and overcoming these challenges. Among various atmospheric conditions, snow presents a uniquely complex challenge. The heterogeneous nature and varying densities of snowflakes significantly impact the performance and reliability of THz communication links [15, 16]. The interaction between THz waves and snowflakes results in scattering effects, the full scope and consequences of which are still being extensively explored and remain largely unknown [17, 18]. This gap in knowledge is further highlighted by the complexities involved in accurately measuring and characterizing the impact of snow on THz wave propagation [19].

This work is dedicated to analyze the complexities of THz channel performance in snowy environments. Conducted on the rooftop of Building 4 at Beijing Institute of Technology, we utilize a comprehensive empirical approach, supported by an array of sophisticated measurement instruments, to thoroughly assess the behavior of THz channels under snowy conditions. This empirical methodology is complemented by theoretical models, offering a holistic and multi-dimensional view of the dynamics of THz channels in such environments. This extensive investigation is crucial for developing a robust model for THz channels, an integral component in the architecture of future wireless networks, particularly with the advent of the anticipated 6G network paradigm.

**II. Channel measurement setup**

The experimental measurement was conducted in an open-air setting, deploying a fixed point-to-point channel configuration as illustrated in Fig. 1(a). This arrangement required precise positioning of the transmitter and receiver, as depicted in Figs. 1(b) and (c), respectively. For the 113-170 GHz channel investigation, the components were positioned 21 meters apart. In contrast, for the 220-325 GHz channel, this distance was reduced to 11 meters, a modification necessitated by its lower transmit power and the increased path loss characteristic of higher frequency transmission. The rooftop of Building 4 at the Beijing Institute of Technology (BIT) was chosen as the experimental site, owing to its conducive environment for atmospheric research.

Central to the transmission system was the Ceyear 1465D signal generator, capable of generating signals up to 20 GHz. These signals were then up-converted to the target THz range (113-170 GHz) using the Ceyear 82406B frequency multiplier module. The Continuous Wave (CW) signals, spanning 113 to 170 GHz, were radiated through a horn antenna (HD-1400SGAH25), complemented by a dielectric lens with a focal length of 10 cm to enhance the signal transmission range. Concurrently, the receiver was equipped with a corresponding horn antenna and lens, directing the incoming signals to a Ceyear 71718 power sensor for direct detection. For the higher frequency range (220-325 GHz), the setup comprised a Ceyear 82406D frequency multiplier paired with a Ceyear 89901S horn antenna. Both transmission and reception units were elevated 80 cm above the floor, exceeding the radius of the first Fresnel zone (9-11 cm for the 113-170 GHz channel and 5-6 cm for the 220-325 GHz channel), and were shielded by a tarp to prevent snow intrusion. The signal beamwidth was measured at 5.7º for 140 GHz, and approximately 4º for 270 GHz. A laptop was employed for control and data acquisition, managing the signal generator and power meter at a data recording frequency of 7 Hz. This equipment setup is in line with previous studies on scattering performance [20, 21] and physical-layer security aspects [22].

To comprehensively assess channel behavior, observations were carried out under two distinct meteorological scenarios: clear weather and snowy conditions. Over a two-day span, data was recorded at 35-second intervals, yielding a substantial dataset representative of each weather scenario. Supplementing this empirical data were atmospheric measurements from the Meteorological Observation Center of the China Meteorological Administration. Data from a meteorological radar situated near the BIT enriched our analysis, providing crucial theoretical insights into link performance amid varying atmospheric dynamics.

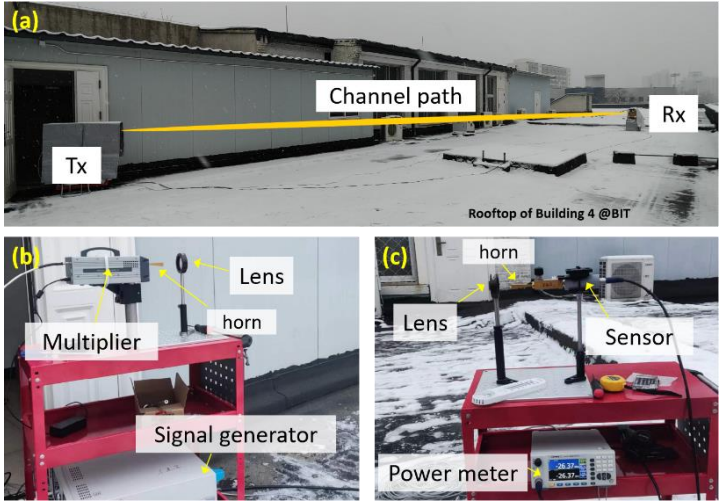

Figure 1 THz channel measurement setup implemented in the campus of Beijing Institute of Technology (BIT). (a) Outdoor channel on the rooftop of Building 4 at BIT with both transmitter (Tx) and receiver (Rx) safeguarded by waterproof coverings; (b) Transmitter hardware; (c) Receiver hardware.

### III. Analysis on channel power profile

Our investigation focused on the efficacy of the outdoor point-to-point THz channel, primarily measured in terms of received power, a critical metric in the design and assessment of wireless communication systems. We concentrated on the 140 GHz and 270 GHz frequencies, where our experimental setup yielded optimal output power. Measurements at 140 GHz were conducted on December 12, 2023, under conditions of approximately -1°C temperature and 52% relative humidity (RH), while those at 270 GHz, carried out on December 13, 2023, recorded a marginally lower ambient temperature of 0°C and higher RH of 60%. These conditions, bordering on the freezing point with elevated humidity, are suggestive of wet snow, a pivotal element in our analysis.

In evaluating THz channel performance under both clear and snowy conditions, we employed the cumulative distribution function (CDF) to scrutinize the power measurements. The data, shown in Figs. 2(a) and (b), was collected at varied intervals, adopting the liquid water equivalent (LWE) precipitation rate as a more precise gauge of snowfall intensity compared to traditional visibility metrics. This methodology, informed by the diversity in snow types and their distinct impacts on visibility [23], is in line with the LWE intensity guidelines set forth by the SAE Ground De-icing committee in 1988 [24]. Our results, detailed in Table I, indicate a discernible trend of power reduction across both the 140 GHz and 270 GHz channels as LWE rates increasing, hinting that extended transmission distances may amplify this loss. This observation underscores the potential necessity for counteracting methodologies and techniques in THz wireless communication systems to counteract the effects of various atmospheric conditions.

A comparison of the CDF curves in clear versus snowy conditions revealed no substantial difference as in Fig. 2, implying minimal scattering effects from snowfall. This was substantiated by fitting the CDF data to a Rician distribution model, as illustrated in Table I. The correspondence of the CDF with a Rician distribution in both conditions, as opposed to the expected Weibull distribution [18], marks a notable deviation. We attribute this to atmospheric turbulence [19] and its influence on pointing errors. It is noteworthy that wind speeds during our measurements were relatively mild, below 5 m/s (light breeze), thus unlikely to induce significant turbulence-induced pointing errors, especially for THz frequency band [5].

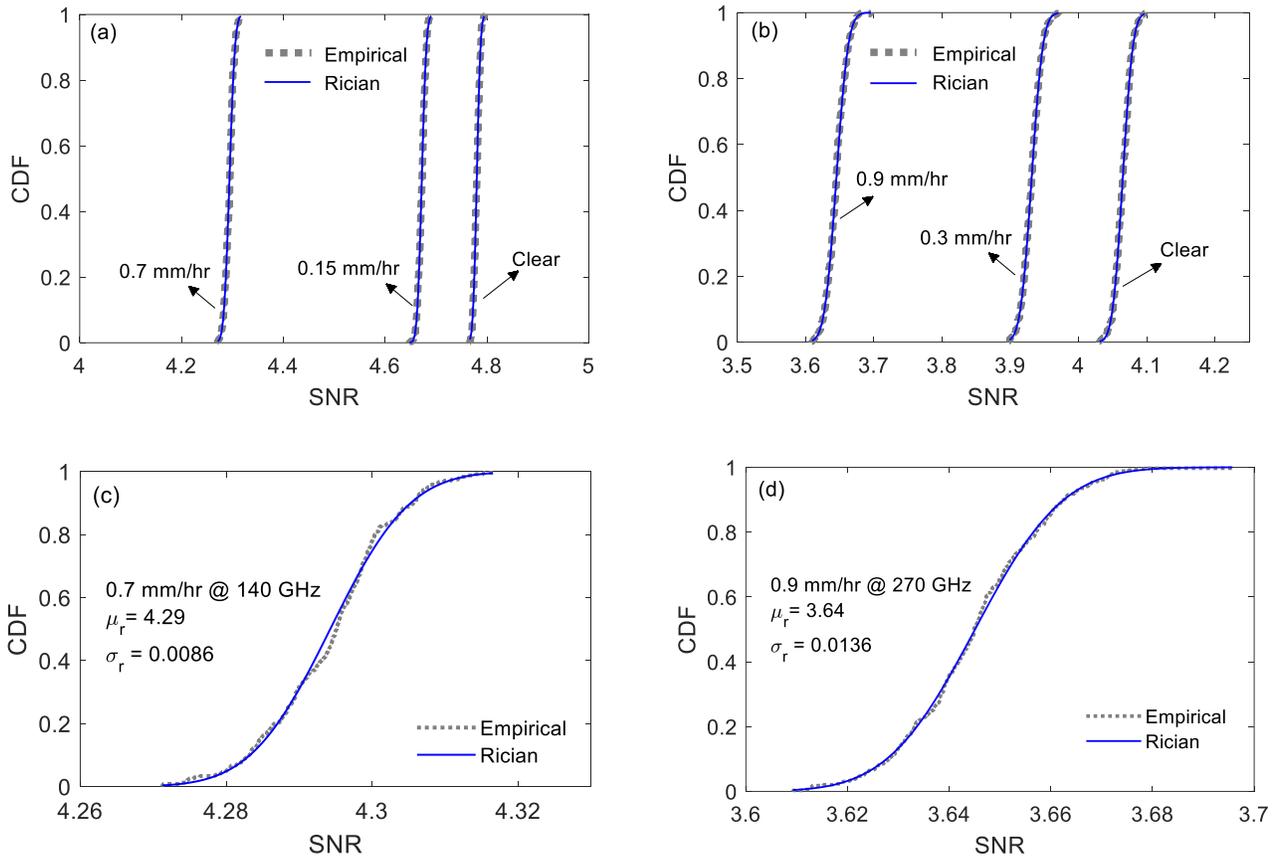

Figure 2 CDF profile for received SNR with and without snowfall with operating frequencies at (a) 140 GHz and (b) 270 GHz, alongside the fitted CDF to the measured data at (c) 140 GHz and (d) 270 GHz in snowfall conditions.

K factor refers to the ratio of the power of the LoS path signal to the power of the reflected or scattered signals (multipath components). The observed high K factor in snowy conditions, as shown in Table I, suggests a dominant LoS component. This leads us to infer that multipath effects, typically a result of scattering, are minimal in our measurement scenario again, aligning with findings in rainy conditions [25]. This is further supported by the observation that the presence of snowflakes did not significantly contribute to multipath components, consistent with prior studies on multipath profiles [15]. Additionally, the increased propagation loss due to the elongated path length of multipath components, relative to the LOS path, also supports this conclusion.

Nevertheless, this does not invalidate the impact of snow on the channel's phase performance. The variability in signal strength, indicated by the variance $\sigma_r^2$ in the received signal's signal-to-noise ratio (SNR), is ascribed to dynamic interactions with moving snowflakes. These interactions result in signal strength fluctuations and illuminate the multiplicative nature of signal degradation over distance, a phenomenon also noted in rainy conditions [8, 26]. In this work, it is crucial to account for the statistical distribution of signal variability along with time-variant channel characteristics, such as rapid phase and amplitude changes. Although we emphasize the need to consider time-variant LWE precipitation rates to fully capture these rapid fluctuations, the limitations in meteorological data acquisition, typically performed at hourly intervals, restrict our ability to thoroughly characterize these fast fluctuations (scintillation effects) observed in atmospheric turbulence measurements [5]. Furthermore, temporal variations in snowfall density and distribution could impact the consistency of signal reception. Nonetheless, our measurement protocol, with a recording interval of 1/7 s, exceeds the transient period (approximately 0.03s) required for a wet snowflake, moving at an average speed of 1.5 m/s [16], to cross the beam width of roughly 5 cm in diameter, posing a challenge in distinctly identifying and analyzing these rapid fluctuations.

Table I Parameters for fitting to the Rician distribution. (Parameter $\mu_r$ represents the mean value of the received SNR, $\sigma^2$ demotes the variance of the received signal's SNR, and $K$ symbolizes the $K$-factor for the Rician distribution across different frequencies and weather conditions).

| Frequency | Weather condition | $\mu_r$ | $\sqrt{\sigma_r^2}$ | $K$ |
|---|---|---|---|---|
| 140 GHz | Clear | 4.78 | 0.0058 | 55.3 dB |
| | Snowy (0.15 mm/hr) | 4.67 | 0.0070 | 53.4 dB |
| | Snowy (0.7 mm/hr) | 4.29 | 0.0086 | 51 dB |
| 270 GHz | Clear | 4.06 | 0.0118 | 47.7 dB |
| | Snowy (0.3 mm/hr) | 3.93 | 0.0130 | 46.6 dB |
| | Snowy (0.9 mm/hr) | 3.64 | 0.0136 | 45.6 dB |

**IV. Analysis on channel power loss**

In this section, we explore the intricate dynamics of snow and its impact on terahertz (THz) signal propagation, with a specific focus on power loss. To effectively quantify the power loss attributable to snow, it is imperative to monitor the temporal variations and aggregate these data to deduce average values and corresponding uncertainties. As demonstrated in Fig. 3, our data unambiguously indicates a progressive increase in power loss correlating with rising snowfall rates, corroborating trends identified in previous studies [17]. However, the challenge lies in accurately modeling these fluctuations, given the diverse forms and distribution of snowflakes, as stated above.

In the analytical approach, we have employed several theoretical models, including ITU-R P.1817-1 [27], the Gunn-East model based on the Rayleigh approximation [28], and the renowned Mie scattering theory [29], in conjunction with the Scott distribution for snowflakes [30]. These models provide a nuanced understanding of the impact of snow on THz signal propagation. The Scott distribution, expressed as a negative exponential function $N(r) = N_0 \exp(-\Lambda r)$, with parameters $N_0 = 100 \times 10^3$ and $\Lambda = 5.76 R^{-0.31}$, proved particularly useful in our calculations, offering a more refined analysis compared to other distributions like Marshall-Palmer, Gunn-

Marshall, and Sekhon-Srivastava. The Mie scattering theory models the dielectric properties of snow using Debye's mixture theory [16, 31], employing the Double-Debye dielectric model (D3M) [32] for water and a single Debye model for ice [33, 34]. Additionally, gaseous absorption along the channel path was accounted for using the ITU-R recommendation model [35], validated for frequencies between 1 and 450 GHz [36].

Our theoretical prediction, as depicted in Fig. 3(a) and (b), reveals that while the ITU model tends to overestimate power loss, the Gunn-East model, more apt for millimeter-wave channels below 70 GHz, significantly underestimates it. The Mie scattering theory (called Scott model following), given the size of snow particles relative to THz wavelengths [37], provides a closer approximation, albeit with occasional underestimations as evidenced in previous findings [17]. The empirical data resides between the predictions by the Scott and ITU models, suggesting that a combined approach integrating both models could yield more accurate predictions. For our computations, wet snow parameters were considered, in line with the near-freezing temperatures and high humidity during our measurements. Interestingly, the ITU model for dry snow more closely matched our empirical data than that for wet snow, prompting its inclusion in the following analyses. The estimated wetness ($m_v$) of the snowflakes, assumed to be 10%, was not empirically measured but estimated.

Addressing the complexities of accurately modeling channel power attenuation due to snowfall, we also considered empirical fitting, as suggested in ITU-R Recommendation. This method, defined by the equation

$$\gamma_{snow} = a \cdot \mathrm{Rr}^b \, [\mathrm{dB/km}] \qquad (1)$$

where $a$ and $b$ are empirically derived constants, and Rr refers the LWE rate of precipitation in mm/hr, offers a practical approach for aligning the model with observed data. For this, we obtained constants $a = 29.79$, $b = 0.82$ for 140 GHz, and $a = 30.28$, $b = 1.15$ for 270 GHz by a power-fitting algorithm. This technique is advantageous in scenarios where theoretical models are insufficient to capture the intricate interactions between THz frequencies and snow particles. However, this specific attenuation model may not be universally applicable across varied environmental conditions or geographical locations, as it is typically tailored to specific datasets or conditions. Another limitation is its reliance on the availability and quality of measured data, without providing deeper theoretical insights. Consequently, we intend to continue employing the Scott model for further theoretical investigations.

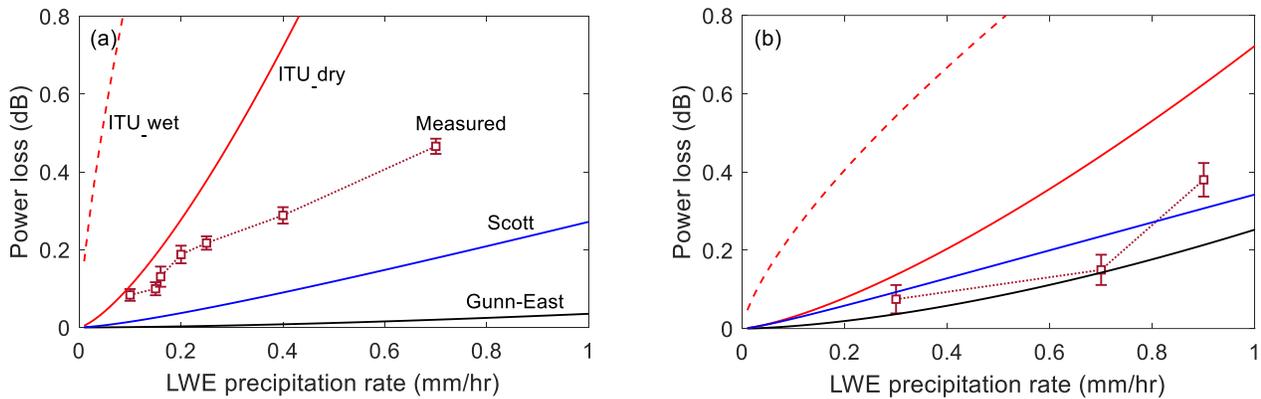

Figure 3 Power loss relative to the LWE precipitation rate for the 140 GHz and 270 GHz channels, over distances of 21 m and 11 m, respectively. (b) keeps an identical legend with (a).

**V. Theoretical analysis on channel performance**

The preceding theoretical model illuminates signal attenuation and power profile in THz channels. However, translating these findings into a comprehensive understanding of communication link reliability requires a deeper exploration. To this end, bit-error-ratio (BER) analysis becomes an indispensable tool, offering critical insights into how power variations influence the error rates in transmitted data. Such an analysis is not only fundamental for evaluating the channel's efficacy but also necessary for designing and optimizing wireless communication systems in the THz band, where atmospheric elements like humidity, precipitation, and temperature profoundly affect signal transmission. By dissecting the impact of these atmospheric factors on BER, it is helpful to refine antenna design, signal processing algorithms, and error correction techniques to reinforce the system's overall performance and reliability.

In this section, we conduct a detailed theoretical examination of channel performance under diverse LWE precipitation rates. As depicted in Fig. 4(a), an obvious observation emerges: the discrepancy between the ITU and Scott model predictions diminishes as the carrier frequency increases, suggesting a heightened consistency and reliability of both models at higher frequency bands. Notably, the Scott model, corroborated by previous research [13], more accurately forecasts an increased power loss at higher frequencies. Concurrently, the ITU model exhibits improved accuracy as the carrier frequency nears the upper band, showing closer alignment with the optical frequency range.

Our efforts focus then shifts to the BER performance of various channels under clear and snowy conditions, applying an LWE precipitation rate of 1 mm/hr. The selected modulation schemes for this analysis are amplitude shift keying (ASK) and 16-quadrature amplitude modulation (16-QAM), chosen based on their deployment in our prior studies [8, 16, 26]. ASK's sensitivity to amplitude fluctuations makes it particularly suitable for assessing the influence of varying atmospheric conditions. Conversely, QAM, utilizing both amplitude and phase changes for data transmission, offers a more comprehensive evaluation of signal integrity under different weather scenarios. The BER performance is predicted using established mathematical expressions,

$$\text{BER}_{ASK} = Q\left(\sqrt{2 \cdot SNR}\right) \quad (2)$$

and

$$\text{BER}_{16-QAM} = \frac{4}{\sqrt{M}} \cdot Q\left(\sqrt{\frac{3 \cdot SNR}{M-1}}\right) \quad (3)$$

incorporating the *Q*-function to denote the probability of a Gaussian random variable exceeding a specified value, with modulation order denoted by *M*. The operational frequencies under consideration - 140 GHz, 220 GHz, and 340 GHz - fall within atmospheric transparency windows and are commonly used in various wireless link configurations [38].

In Figs. 4(b), (c), and (d), the upper and lower bounds of the predicting area correspond to the ITU and Scott models, respectively. A predictable finding is that lower carrier frequencies and simpler modulation orders, such as ASK, lead to improved BER performance. This enhancement at lower frequencies is ascribed to the greater power loss typically observed at higher frequencies. Moreover, advanced modulation schemes necessitate a higher SNR for equivalent BER values, rendering them more prone to errors in snowy conditions. It is noteworthy that under specific conditions, error-free data transmission (BER<$10^{-10}$) is feasible, particularly at lower frequencies and with simpler modulation techniques. This underscores the strategic selection of frequency and modulation strategy for THz communications in challenging atmospheric scenarios.

Optimal THz channel performance is observed with LWE precipitation rates below 1 mm/hr, typically associated with light snowfall conditions [24]. This finding indicates that light snowfall exerts minimal impact on THz communication over distances of up to 1 km. Nevertheless, the reliability of the channel decreases with either an increase in distance or snowfall intensity. Incorporating forward error correction (FEC) may could enhance error resilience by correcting errors within its threshold (typically $2\times10^{-3}$), although its effectiveness is bounded by the complexity of the coding scheme and the level of redundancy introduced.

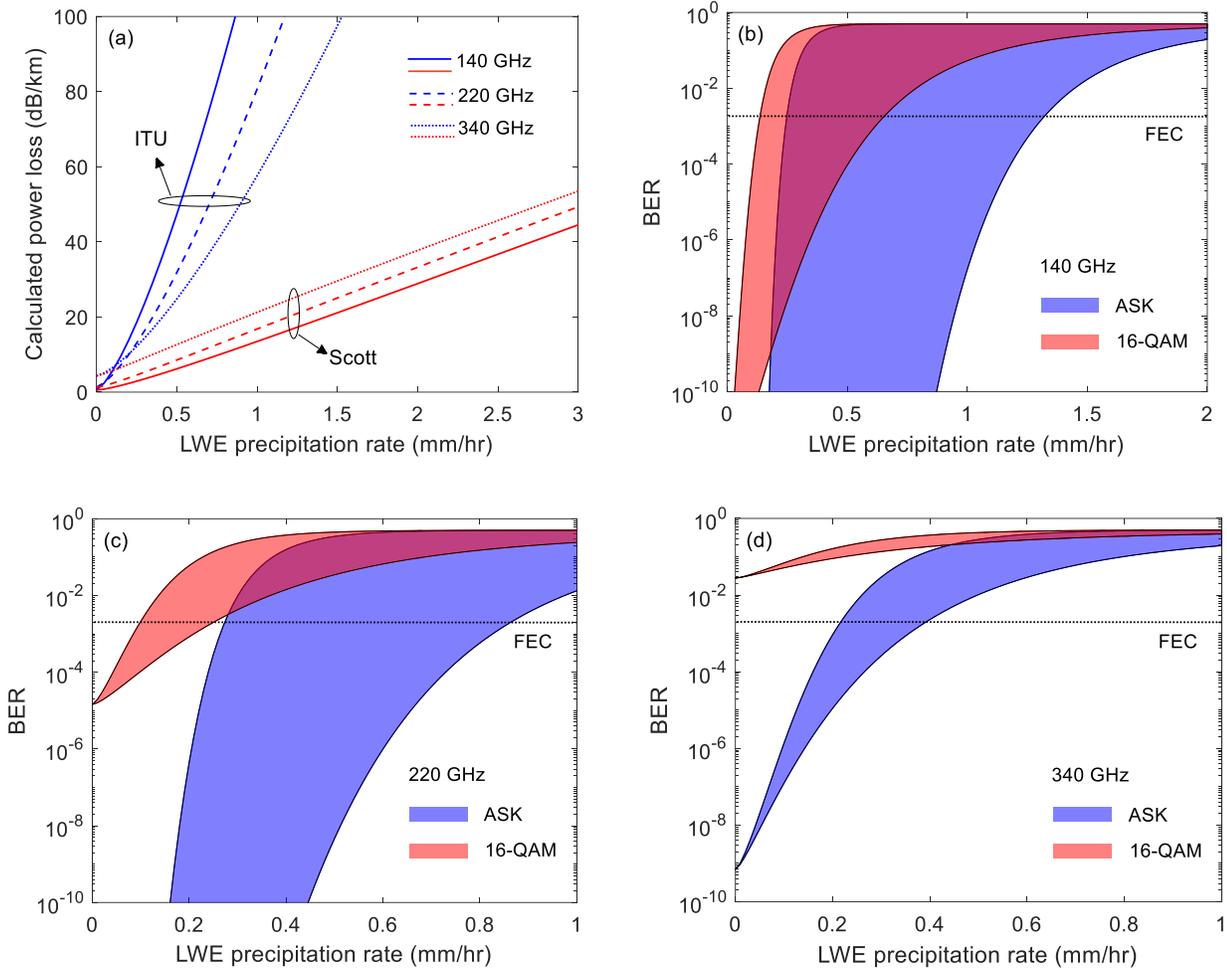

Figure 4 (a) Calculated average power loss with respect to LWE rate suffered by channels operating at 140, 220 and 340 GHz. Predicted average BER performance of the channels operating at (b) 140 GHz, (c) 220 GHz and (4) 340 GHz. The upper and lower bounds of the predicting area correspond to the predictions by ITU model and Scott model, respectively. Channel distance 1 km, relative humidity RH 60%, temperature 0°C, transmitted power 20 dBm, noise level of receiver -60 dBm; the gain at the transmitter and receiver side are identical and equals to 40 dB (combination of antenna and lens).

To broaden our understanding of channel performance in relation to operating frequency, we conducted an extensive analysis encompassing power loss and BER evolution. This analysis, accounting for atmospheric absorption factors such as 60% RH and a temperature of 0°C, utilized the Scott model. Setting the LWE precipitation rate at 0.5 mm/hr and maintaining the channel distance at 1 km, we observed in Fig. 5(a) that lower frequencies exhibit reduced sensitivity to changes in humidity. Under ASK modulation, the operational frequency can reach up to 420 GHz in clear conditions, while it is limited to 260 GHz under 16-QAM in similar conditions.

However, in snowy scenarios, the threshold frequency drops to 260 GHz for ASK and 140 GHz for 16-QAM, considering the FEC threshold. Notably, variations in atmospheric humidity also influence the wetness of snowflakes, a factor not accounted for in this theoretical model due to the complex mechanisms underlying this interaction.

As ambient temperature increases, the air's capacity to retain water vapor (saturation vapor density) also rises. Consequently, if vapor density remains constant while temperature escalates, relative humidity decreases due to the air's enhanced capacity to hold more water vapor [39]. In the context of THz communications, especially under conditions of wet snowfall (temperatures ranging from 0°C to 3°C), understanding the interplay between atmospheric conditions and THz wave propagation is crucial. Our findings in Fig. 5(b) suggest that within this specific temperature range, temperature variations have negligible influence on THz frequency propagation below 200 GHz. This implies that despite potential changes in temperature affecting relative humidity, the propagation characteristics of THz channel within this frequency band largely remain stable, a critical factor for ensuring consistent communication quality and reliability in regions experiencing wet snowfall.

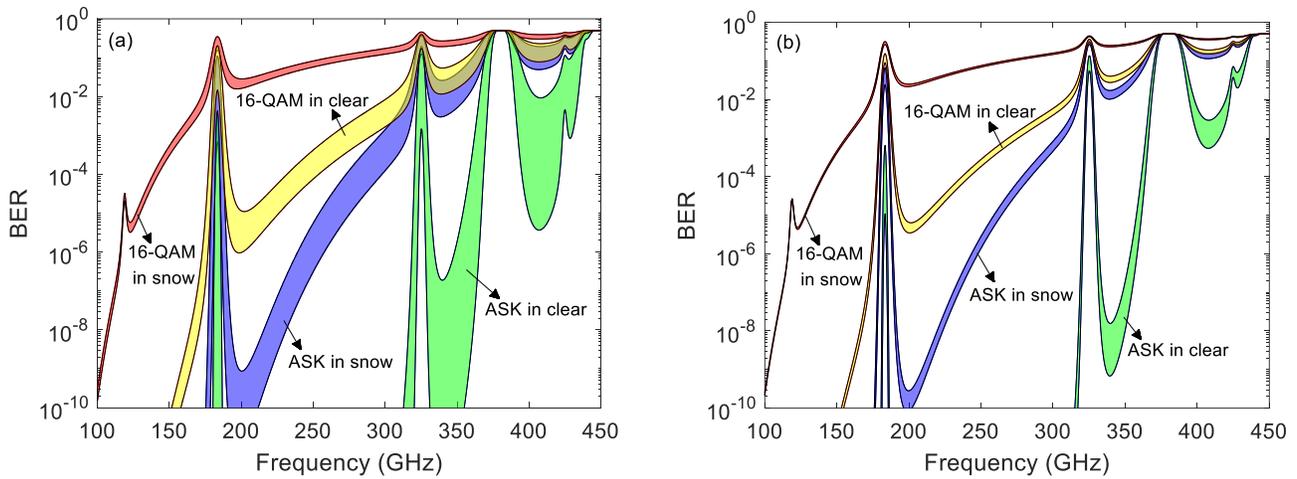

Figure 5 Calculated average BER performance due to the variation of (a) relative humidity and (b) ambient temperature. Channel distance 1 km, humidity RH 60%, temperature 0°C, transmitted power 20 dBm, noise level of receiver -60 dBm, LWE rate 0.5 mm/hr; the gain at the transmitter and receiver side are identical and equals to 40 dB (combination of antenna and lens).

## VI. Conclusions

In this work, we conducted a comprehensive investigation on terahertz (THz) channel dynamics under snowy conditions, utilizing a combination of outdoor measurements and advanced channel modeling techniques. A specialized measurement setup was established at the BIT, facilitating an in-depth analysis on the power profile, link budget, and BER performance in snowy weather conditions. Our findings were particularly revealing in the context of CDF profiles under both clear and snowy conditions. These profiles consistently conformed to Rician distributions with variable $K$-factors, an observation pointing towards the minimal impact of multipath effects in our experimental setup. Additionally, our study highlighted the rapid power fluctuations induced by the dynamic interaction between THz waves and moving snowflakes.

In the realm of theoretical modeling, the Scott model emerged as a more accurate predictor. Employing it for BER predictions revealed that lower carrier frequencies and simpler modulation orders, lead to improved BER performance, even though light snowfall can significantly impact BER performance. This work also delved into the influence of ambient humidity and temperature on THz communication, finding that lower frequencies are less susceptible to humidity variations and that within a specific temperature range, the impact of temperature changes on THz frequency propagation below 200 GHz is negligible.

Despite these contributions, this work faced certain limitations. The current experimental setup's inability to accurately capture rapid power fluctuations limited our investigation into scintillation effects, an area we aim to focus on in future. Additionally, we hypothesize that snowflakes, when moving at sufficient velocities, might induce a Doppler shift in the received THz signal, potentially affecting the signal's phase. This phenomenon is likely to be more pronounced at higher transmission frequencies and thus merits further investigation.

**Acknowledgement**

This work was supported in part by the National Natural Science Foundation of China (62071046), the Science and Technology Innovation Program of Beijing Institute of Technology (2022CX01023), the Talent Support Program of Beijing Institute of Technology "Special Young Scholars" (3050011182153).

**References**

[1] T. Kürner, D. M. Mittleman, and T. Nagatsuma, *THz Communications: Paving the Way Towards Wireless Tbps*. Cham, Switzerland: Springer, 2021.
[2] G. Liu *et al.*, "Myelin Sheath as a Dielectric Waveguide for Signal Propagation in the Mid-Infrared to Terahertz Spectral Range," *Advanced Functional Materials,* vol. 29, p. 1807862, 2019.
[3] J. Ma, M. Weidenbach, R. Guo, M. Koch, and D. M. Mittleman, "Communications with THz Waves: Switching Data Between Two Waveguides," *Journal of Infrared, Millimeter, and Terahertz Waves,* vol. 38, no. 11, pp. 1316-1320, 2017, doi: 10.1007/s10762-017-0428-4.
[4] Y. Yang, M. Mandehgar, and D. R. Grischkowsky, "Understanding THz Pulse Propagation in the Atmosphere," *IEEE Transactions on Terahertz Science and Technology,* vol. 2, no. 4, pp. 406-415, 2012, doi: 10.1109/tthz.2012.2203429.
[5] J. Ma, L. Moeller, and J. F. Federici, "Experimental Comparison of Terahertz and Infrared Signaling in Controlled Atmospheric Turbulence," (in English), *Journal of Infrared, Millimeter and Terahertz Waves,* vol. 36, no. 2, pp. 130-143, Feb 2015. [Online]. Available: <Go to ISI>://WOS:000349632100003.
[6] J. Ma, R. Shrestha, L. Moeller, and D. M. Mittleman, "Invited Article: Channel performance for indoor and outdoor terahertz wireless links," *APL Photonics,* vol. 3, no. 5, p. 051601, 2018, doi: 10.1063/1.5014037.
[7] Y. Yang, M. Mandehgar, and D. R. Grischkowsky, "Broadband THz Signals Propagate Through Dense Fog," *IEEE Photonics Technology Letters,* vol. 27, no. 4, pp. 383-386, 2015, doi: 10.1109/lpt.2014.2375795.
[8] J. Ma, F. Vorrius, L. Lamb, L. Moeller, and J. F. Federici, "Experimental Comparison of Terahertz and Infrared Signaling in Laboratory-Controlled Rain," (in English), *Journal of Infrared, Millimeter and Terahertz Waves,* vol. 36, no. 9, pp. 856-865, Sep 2015, doi: 10.1007/s10762-015-0183-3.
[9] K. Su, L. Moeller, R. B. Barat, and J. F. Federici, "Experimental comparison of terahertz and infrared data signal attenuation in dust clouds," (in English), *J Opt Soc Am A,* vol. 29, no. 11, pp. 2360-2366, Nov 2012. [Online]. Available: <Go to ISI>://WOS:000310590500014.
[10] K. Su, L. Moeller, R. B. Barat, and J. F. Federici, "Experimental comparison of performance degradation from terahertz and infrared wireless links in fog," (in English), *J Opt Soc Am A,* vol. 29, no. 2, pp. 179-184, Feb 2012. [Online]. Available: <Go to ISI>://WOS:000300396500049.
[11] Q. Jing, D. Liu, and J. Tong, "Study on the Scattering Effect of Terahertz Waves in Near-Surface Atmosphere," *IEEE Access,* vol. 6, pp. 49007-49018, 2018, doi: 10.1109/access.2018.2864102.

[12]   G. A. Siles, J. M. Riera, and P. Garcia-del-Pino, "Atmospheric Attenuation in Wireless Communication Systems at Millimeter and THz Frequencies," *IEEE Antennas and Propagation Magazine,* vol. 57, no. 1, pp. 48-61, 2015, doi: 10.1109/map.2015.2401796.

[13]   E.-B. Moon, T.-I. Jeon, and D. R. Grischkowsky, "Long-Path THz-TDS Atmospheric Measurements Between Buildings," *IEEE Transactions on Terahertz Science and Technology,* vol. 5, no. 5, pp. 742-750, 2015, doi: 10.1109/tthz.2015.2443491.

[14]   G. B. Wu, Y.-S. Zeng, K. F. Chan, S.-W. Qu, and C. H. Chan, "High-gain circularly polarized lens antenna for terahertz applications," *IEEE Antennas and Wireless Propagation Letters,* vol. 18, no. 5, pp. 921-925, 2019.

[15]   P. Sen *et al.*, "Terahertz communications can work in rain and snow: Impact of adverse weather conditions on channels at 140 GHz," presented at the Proceedings of the 6th ACM Workshop on Millimeter-Wave and Terahertz Networks and Sensing Systems, 2022.

[16]   J. Ma, J. Adelberg, R. Shrestha, L. Moeller, and D. M. Mittleman, "The Effect of Snow on a Terahertz Wireless Data Link," *Journal of Infrared, Millimeter and Terahertz Waves,* vol. 39, no. 6, pp. 505-508, 2018, doi: 10.1007/s10762-018-0486-2.

[17]   Y. Amarasinghe, W. Zhang, R. Zhang, D. M. Mittleman, and J. Ma, "Scattering of Terahertz Waves by Snow," *Journal of Infrared, Millimeter, and Terahertz Waves,* vol. 41, pp. 215-224, 2019.

[18]   M. Taherkhani, Z. G. Kashani, and R. Sadeghzadeh, "Average bit error rate and channel capacity of terahertz wireless line-of-sight links with pointing errors under combined effects of turbulence and snow," *Appl Optics,* vol. 59, no. 33, p. 10345, 2020, doi: 10.1364/ao.403390.

[19]   M. Taherkhani, Z. G. Kashani, and R. A. Sadeghzadeh, "On the performance of THz wireless LOS links through random turbulence channels," *Nano Communication Networks,* vol. 23, p. 100282, 2020, doi: 10.1016/j.nancom.2020.100282.

[20]   P. Li *et al.*, "Scattering and Eavesdropping in Terahertz Wireless Link by Wavy Surfaces," *IEEE Transactions on Antennas and Propagation,* vol. Early Access, pp. 1-1, 2023, doi: 10.1109/TAP.2023.3241333.

[21]   J. Liu, P. Li, D. Li, G. Liu, H. Sun, and J. Ma, "High-Speed Surface Roughness Recognition by Scattering on Terahertz Waves," presented at the 16th UK-Europe-China Workshop on Millimetre Waves and Terahertz Technologies (UCMMT), Guangzhou, China, 2023.

[22]   Y. Qiao *et al.*, "Tubular Eavesdropper on a Terahertz Link - A Tentative Study," presented at the 16th UK-Europe-China Workshop on Millimetre Waves and Terahertz Technologies (UCMMT), Guangzhou, China, 2023.

[23]   R. M. Rasmussen, J. Vivekanandan, J. Cole, B. Myers, and C. Masters, "The Estimation of Snowfall Rate Using Visibility," *Journal of Applied Meteorology,,* vol. 38, no. 10, pp. 1542-1563, 1998.

[24]   T. Fahey, "Snowfall Rate Thresholds for Light, Moderate and Heavy," in "Aerodrome Meteorological Observation and Forecast Study Group (AMOFSG), Seventh Meeting," Montréal, AMOFSG/7-IP/4. 27/.6/08, 9 to 12 September 2008.

[25]   R. Wang, Y. Mei, X. Meng, and J. Ma, "Secrecy Performance of Terahertz Wireless Links in Rain and Snow," *Nano Communication Networks,* vol. 28, p. 100350, 2021, doi: 10.1016/j.nancom.2021.100350.

[26]   P. Li *et al.*, "Performance degradation of terahertz channels in emulated rain," *Nano Communication Networks,* vol. 35, p. 100431, 2023.

[27]   "Recommendation ITU-R P.1817-1: Propagation data required for the design of terrestrial free-space optical links."

[28]   K. L. S. GUM and T. W. R. East, "The microwave properties of precipitation particles," *The Quarterly Journal of the Royal Meteorological Society,* vol. 80, no. 346, pp. 522-545, 1954.

[29]   D. Deirmendjian, *Electromagnetic scattering on spherical polydispersions*. New York: American Elsevier Publishing, 1969.

[30]   B. C. Scott, "Theoretical estimates of the scavenging coefficient for soluble aerosol particles as a function of precipitation type, rate and altitude," *Atmospheric Environment,* vol. 16, no. 7, pp. 1753-1762, 1982.

[31]   J. E. Jiusto, "Types of snowfall," vol. 54, no. 11, pp. 1148-1162, 1973.

[32]  F. T. Ulaby, R. K. Moore, and A. K. Fung., *Microwave Remote Sensing: Actice and Passive. Volume II-Radar Remote Sensing and Surface Scattering and Emission Theory*. Norwood, Massachusetts: Artech House, 1982.

[33]  O. Mishima, D. D. Klug, and E. Whalley, "The far-infrared spectrum of ice Ih in the range 8–25 cm−1. Sound waves and difference bands, with application to Saturn's rings.," *J Chem Phys,* vol. 78, no. 11, pp. 6399-6404, 1983.

[34]  J. H. Jiang and D. L. Wu, "Ice and water permittivities for millimeter and sub-millimeter remote sensing applications," *Atmospheric Science Letters,* vol. 5, no. 7, pp. 146-151, 2004, doi: 10.1002/asl.77.

[35]  "International Telecommunication Union Recommendation (ITU-R) P.676-13: Attenuation by atmospheric gases and related effects." https://www.itu.int/rec/R-REC-P.676-13-202208-I/en (accessed.

[36]  J. F. Ohara and D. R. Grischkowsky, "Comment on the Veracity of the ITU-R Recommendation for Atmospheric Attenuation at Terahertz Frequencies," *IEEE Transactions on Terahertz Science and Technology,* vol. 8, no. 3, pp. 372-375, 2018, doi: 10.1109/tthz.2018.2814343.

[37]  H. R. Pruppacher and J. D. Klett, *Microphysics of clouds and precipitation*. Dordrecht: Kluwer Academic Publishers, 1997.

[38]  T. Nagatsuma, G. Ducournau, and C. C. Renaud, "Advances in terahertz communications accelerated by photonics," *Nature Photonics,* vol. 10, no. 6, pp. 371-379, 2016, doi: 10.1038/nphoton.2016.65.

[39]  C. D. Ahrens, *Meteorology today: an introduction to weather, climate, and the environment*. Virginia, U.S.: Thomson/Brooks/Cole, 2003.